\documentclass[a4paper]{article}
\usepackage{qcircuit}
\usepackage{graphicx}
\usepackage[utf8x]{inputenc}
\usepackage[T1]{fontenc}
\usepackage{authblk}
\usepackage{subfig}
\usepackage[a4paper,top=3cm,bottom=2cm,left=3cm,right=3cm,marginparwidth=1.75cm]{geometry}
\input{amssym}
\usepackage[colorinlistoftodos]{todonotes}
\usepackage[colorlinks=true, allcolors=blue]{hyperref}
\usepackage{authblk}

\date{}

\title{\bf Investigation of Optical Pumping in Cesium
atoms with a Radio-Frequency field, Using Liouville equation}
\author[1]{H. Davoodi Yeganeh\thanks{h.yeganeh@ut.ac.ir }}
\author[1]{Z. Shaterzadeh-Yazdi\thanks{zahra.shaterzadeh@ut.ac.ir }}

\affil[1]{School of Engineering Science, College of Engineering, University of Tehran, Tehran, Iran, 143-9955961}

\begin{document}
\maketitle

\begin{abstract}
Optical pumping is a powerful technique used to manipulate the population of atomic sublevels in specific atoms. In this study, we focus on investigating the population dynamics of Cesium atoms using the Liouville equation. By applying circularly polarized light at an appropriate frequency for electronic transitions from ground to excited states, we analyze the relaxation rate, repopulation, and evolution of Cesium Zeeman sublevels. To further control the sublevel population post-optical pumping, we introduce a radiofrequency (RF) field into the Liouville equation calculations. Through this integrated approach, we successfully achieve a tailored distribution of atomic sublevel populations with high efficiency. This method holds promise for various applications in optical experiments.
\end{abstract}

\section{Introduction}
Optical pumping is a phenomenon where photons interact with the constituent atoms of a material. In an isolated gas of atoms, the atoms occupy energy states at a given temperature following standard statistical mechanics principles. When these atoms are exposed to photons, the photons play a crucial role in redistributing the states occupied by the atoms~\cite{r1,r2}. Among various methods utilized for manipulating atomic-sublevel populations, such as chirped laser pulse population transfer~\cite{r3} and laser-induced population transfer~\cite{r4}, optical pumping stands out as a particularly practical approach. This method finds application in diverse fields including magneto-optical trapping and laser cooling~\cite{r5}.

The evolution of atomic sub-level states due to optical pumping can be elucidated by examining the evolution of their density matrix through the application of the Liouville equation. In this framework, the system under consideration is typically treated as a closed system, neglecting interactions between the atomic system and its surrounding environment. However, in practice, the interaction between the atomic system and the environment can often be represented phenomenologically as repopulation and relaxation processes within the atomic system, without explicitly accounting for the environment's presence.

Various methods exist for describing the population evolution of sublevel states during the optical pumping process. One such method is the utilization of rate equations~\cite{r6}, which simplifies by neglecting fast-transient processes in the excited states. Another approach involves employing the Lindblad master equation to describe optical pumping within the context of an open quantum system~\cite{r7}. Alkali atoms are commonly chosen for optical pumping experiments due to their relatively simple atomic structures compared to other elements~\cite{r8,r9,r10,r11,r12,r13}. Alkali atoms such as $^{7}$Li, $^{23}$Na, $^{39}$K, $^{87}$Rb, and$^{133}$Cs typically possess spin-half nuclei and exhibit hyperfine-doublet ground states $n\   ^{2}$S$_{1/2}$ defined by $F=I-J,\ I+J$, along with fine-doublet excited states $n\   ^{2}$P${1/2,\ 3/2}$. These excited states further display hyperfine structures corresponding to $F=I-J\  ...\  I+J$, resulting in two hyperfine levels for $J=1/2$ and four hyperfine levels for $J=3/2$.

In this study, we focus on investigating the optical pumping of $^{133}$Cs atoms using polarized light and manipulating their Zeeman sublevel populations through the application of the Liouville equation. We adopt a semiclassical approach to describe the interaction between optical light and Cs atoms. Initially, Cs atoms are excited to their respective excited states through optical pumping, followed by the introduction of a radiofrequency (RF) field to engineer the sublevel populations. The manipulation of atomic states is achieved through a combination of appropriate laser fields, magnetic fields, and RF radiation~\cite{r18,r16,r15,r14,r17}. Our results demonstrate that employing the RF field facilitates the return of excited sublevel populations to the ground states, resulting in an even distribution of population among the sublevel ground states.

The structure of this paper is as follows: in Section~\ref{sec2}, we present a theoretical model for the optical pumping process using the Liouville equation approach. Section~\ref{sec3} discusses the results obtained from calculating the population evolution of Cesium (Cs) sublevels in the presence and absence of the RF field during optical pumping. Finally, in Section~\ref{sec4}, we provide concluding remarks on our findings.

\section{Theoretical Model of Optical Pumping  with Liouville Equation Approach}\label{sec2}

The evolution of an atomic system over time is succinctly captured by its density matrix. This evolution is primarily influenced by intrinsic properties of the system, the atomic structure, and external fields like static electric and magnetic fields, as well as optical electromagnetic fields. Typically, atomic systems are not completely isolated from their surrounding environments, necessitating the inclusion of phenomenological terms in the evolution equations to model these interactions. Such interactions often manifest as relaxation and repopulation phenomena, including processes like radiative decay and collisions~\cite{r5,r19,r20}.

Our objective is to develop a model for a system comprising a thermally distributed ensemble of atoms with varying velocities confined in a vapor cell. The time evolution of the density matrix $\rho$ associated with the system is governed by the equation~\cite{r20}:
\begin{equation}\label{eq1}
i \frac{d}{dt}\rho=[H,\rho]-i \frac{1}{2}{\Gamma, \rho}+i \Lambda,
\end{equation}
where $H$ represents the total Hamiltonian of the system, defined as $H=H_0+H_I+H_B$. Here, $H_0$ corresponds to the Hamiltonian of the target atoms, $H_I$ denotes the light-atom interaction Hamiltonian, and $H_B$ accounts for the magnetic field-atom interaction. The parameter $\Gamma$ signifies the diagonal-form relaxation matrix, reflecting the impact of relaxation on the density matrix's time evolution. In our atomic system of interest, each basis state $|n\rangle$ relaxes at a rate of $\Gamma_n$. The condition $Tr(\rho)=1$ ensures conservation of the number of atoms in the system, necessitating repopulation to replenish atoms. Repopulation is characterized by the repopulation matrix $\Lambda$ in Eq.~\ref{eq1}.

In this study, we meticulously consider all essential aspects of optical pumping, leveraging polarized light for light-matter interactions and adhering to selection rules governing atomic transitions. The interaction Hamiltonian, $H_I$, incorporates interactions between the atomic system and static electric and magnetic fields, along with a radiofrequency field. The entire spectrum of states within the atomic system is comprehensively captured using the associated density matrix. By incorporating selection rules, we can determine the rates of repopulation and relaxation within the target system.

It is noteworthy that in our utilization of the Liouville approach for modeling optical pumping, the inclusion of external fields like RF fields into the Hamiltonian is seamlessly achieved. This contrasts with other methods such as rate equations, which present mathematical limitations in incorporating external fields. Consequently, we opt for the Liouville equation to study optical pumping in Cesium atoms. Through the derivation of the density matrix's time evolution, we gain deeper insights into the population dynamics and coherent transitions within the target atomic system. This method enables us to effectively model optical pumping, manipulate system states, and efficiently utilize optical pumping in alkali atoms for various applications.

\section{Results: Population evolution of the Cesium sublevels }\label{sec3}

\begin{figure}[h!]
\centering
\includegraphics[width=11.5cm]{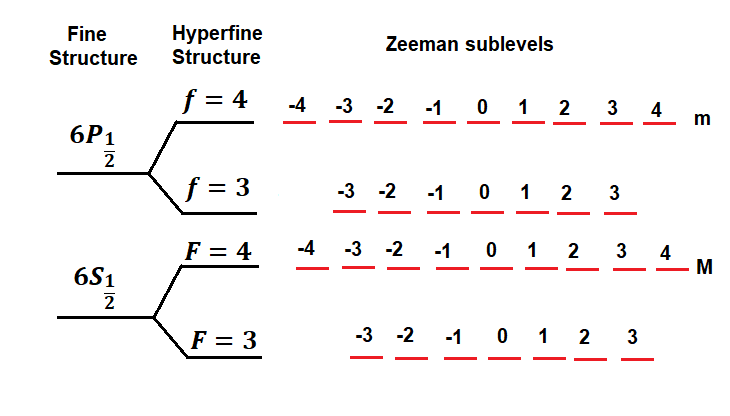}
\caption{Schematic view of fine structure, hyperfine structure and Zeeman splitting of D1 line of Cs atom, employed in  optical pumping
process.}\label{f1}
\end{figure}

Cesium atoms have been extensively utilized in various quantum optics experiments, including the manipulation of cold atoms through laser cooling and trapping~\cite{r22,r21}. In Cs atoms, the fine-structure doublet encompasses two key transitions: $6 ^2 S_{1/2}\rightarrow 6 ^2 P_{3/2}$ and $6 ^2 S_{1/2}\rightarrow 6 ^2 P_{1/2}$, each exhibiting additional hyperfine structures. The ground state of Cs features $J=1/2$ and $I=7/2$, resulting in two hyperfine groundstate levels ($F=3,4$) and two excited hyperfine levels ($f=3,4$) within the D1 transition line. Moreover, each of these hyperfine levels comprises $2F+1$ Zeeman sublevels~\cite{r23}. The schematic representation of the fine structure, hyperfine structure, and Zeeman splitting of D1 lines in Cs atoms is depicted in Fig.~\ref{f1}.

Cesium atoms possess 16 ground-state sublevels, denoted as 
\begin{equation}
|F=3,M=3\rangle, \ldots, |3,-3\rangle, \ldots|4,4\rangle, \ldots, |4,-4\rangle,
\end{equation}
and 16 excited-state sublevels, represented by
\begin{equation}
|f=3,m=3\rangle, \ldots, |3,-3\rangle, \ldots|4,4\rangle, \ldots, |4,-4\rangle.
\end{equation}
The states ${|F,M\rangle}$ and ${|f,m\rangle}$ serve as the basis states for the system's Hilbert space, leading to the representation of the Cs atom's density matrix as a $32\times 32$ matrix.

Considering the ground-state sublevels' energy as zero and the excited-state sublevels' energy as $\hbar\omega_0$, the matrix $H_0$ comprises zero elements except for the diagonal elements representing the excited-state sublevels, i.e.
\begin{equation}
|f,m\rangle \langle f,m|=|4,4\rangle \langle 4,4|= \ldots=|3,-3\rangle \langle 3,-3| =\hbar\omega_0.
\end{equation}
Furthermore, the interaction between Cs atoms and polarized light involves both right-circularly polarized light ($\sigma^+$) and left-circularly polarized light ($\sigma^-$). Therefore, the light-atom interaction Hamiltonian is given by:
\begin{equation}
H_I =-\bf{E}.\bf{\hat d},
\end{equation}
where $\bf{E}$ is the optical electric field and  $\bf{\hat d}$ is the dipole operator representing the electric dipole moment corresponding to the Cs atom.

The electric fields associated with $\sigma^+$ and $\sigma^-$ are assumed to be $E^+=(E_0 e^{i\omega t},i E_0 e^{i\omega t},0)$ and $E^-=(E_0 e^{i\omega t},-i E_0 e^{i\omega t},0)$, respectively. Additionally, we define $d_x= \frac{1}{\sqrt{2}}(d_{-1}-d_{+1})$, where $d_{-1}$ and $d_{+1}$ represent dipole operator matrix elements for $\sigma^+$ and $\sigma^-$ light. These matrix elements are determined using the Wigner–Eckart theorem and are given by,
\begin{equation}
\langle F_1 m_1|d_{\pm}|F_2m_2\rangle=(-1)^{F_1-m_1}\langle F_1 m_1|d|F_2m_2\rangle \pmatrix{ F_1 & 1& F_2 \cr -m_1 & \pm 1 & m_2},
\end{equation}
where $\pmatrix{ F_1 & 1& F_2 \cr -m_1 & \pm 1 & m_2}$  is Clebsch-Gordan coefficients describing how individual angular momentum states may be coupled to yield the total angular momentum state of a system; these coefficients are also known as  3j symbol coefficients. 

\begin{figure}[h!]
\centering
\includegraphics[width=11.5cm]{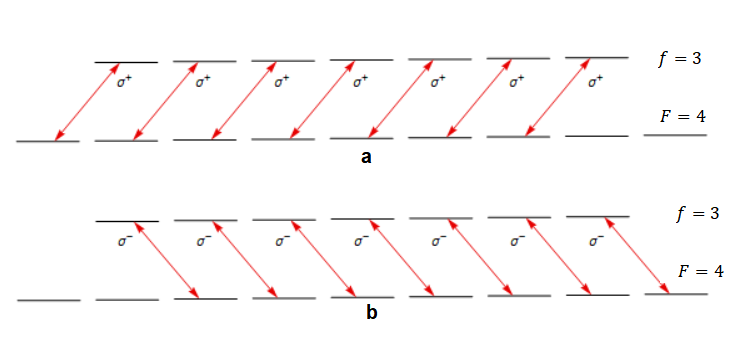}
\caption{Schematic view of transitions between sublevels $F=4$ and $f=3$ with polarized light $\sigma^+$  and (b) polarized light $\sigma^-$ in the D1 line of Cs atoms, which is employed in optical pumping processes.}\label{f3}
\end{figure}

Our focus lies on investigating optical pumping between $F=4$ and $f=3$ in Cs atoms, a process widely employed in quantum memories and other practical applications~\cite{r25}. The interaction between polarized light and these specific sublevels is crucial for various experiments. The Hamiltonian associated with the right-circularly polarized light ($\sigma^+$) interaction is expressed as:
\begin{equation}\label{E55}
\hat{H}_I^+ =-\textbf{E}.\hat{\textbf{d}}=\Omega_R \cos(\omega t) M^+,
\end{equation}
where $\Omega_R$ represents the optical Rabi frequency given by,
\begin{equation}\label{E6}
\Omega_R= E_0 \langle F||d||F'\rangle\pmatrix{ F & 1& F' \cr -m_1 & \pm 1 & m_2}.
\end{equation}
The matrix $M^+$, in Eq.~\ref{E55}, is a $32\times 32$ matrix with non-zero elements only between sublevels $F=4$ and $f=3$, in accordance with selection rules. A similar formulation is applied for the left-circularly polarized light ($\sigma^-$) interaction, i.e.
\begin{equation}
\hat{H}_I^- =-\textbf{E}.\hat\textbf{d}=\Omega_R cos\omega t\  M^-.
\end{equation}
The transitions between sublevels $F=4$ and $f=3$ with $\sigma^+$ and $\sigma^-$ are visually depicted in  Fig.~\ref{f3}. 

Finally, we proceed to introduce the $z$ component of a magnetic field to induce the Zeeman effect. The corresponding interaction Hamiltonian is defined as:
\begin{equation}
\hat{H}_B=\hat \mu .\textbf{B}=g\mu_0F_z B_z=\hbar\Omega_L F_z,
\end{equation}
where $g$ denotes the $g$-factor, $\mu$ represents the magnetic dipole moment, $\Omega_L=g\mu_0 B_z/ \hbar$ signifies the Larmor frequency, and $F_z$ stands for the total angular momentum matrix along the $z$ direction.

Having established the constituent terms of the total Hamiltonian ($H=H_0+H_I+H_B$), we proceed to solve Eq.~\ref{eq1} to explore the dynamics of the system under investigation. Notably, the Hamiltonian exhibits time dependence at the optical frequency. Employing the rotating wave approximation, we disregard rapidly oscillating terms in the Hamiltonian and retain only those reflecting the detuning frequency, $\Delta =\omega-\omega_0$.

Incorporating relaxation at a rate $\gamma$ and spontaneous decay at a rate $\Gamma_s$, the relaxation matrix takes the form of a diagonal $32\times 32$ matrix. The initial 16 diagonal elements correspond to $\gamma$, while the subsequent 16 diagonal elements are characterized by $\gamma +\Gamma_s$. The repopulation matrix $\Lambda$ in Eq.~\ref{eq1} is represented by a $32\times 32$ matrix, encapsulating scenarios where atoms exit the region of interest, potentially replaced by incoming atoms that may exhibit varying degrees of polarization. Additionally, within the repopulation matrix, transition rates between different pairs of upper- and lower-state sublevels are considered, along with coherences between these sublevels that can be influenced by spontaneous decay processes.

Through numerical computation of Eq.~\ref{eq1}, we derive the time evolution of the system's density matrix. Subsequently, by evaluating $Tr(\rho |F,M\rangle\langle F,M|)$, we extract information regarding the population distribution among Zeeman sublevels. It is pertinent to note that the rotating wave approximation employed does not impact the population distribution among Zeeman sublevels in our computational analyses.

\begin{figure}[h!]
\centering
\includegraphics[width=9.6cm]{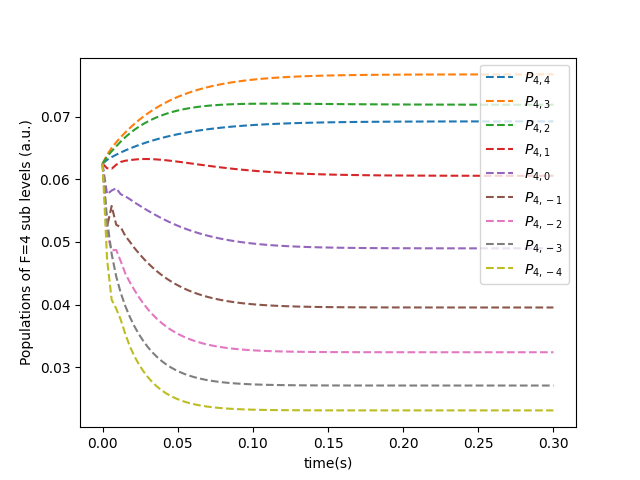}
\caption{Time evolution of the population for the Zeeman sublevels $F=4$. The applied polarized light, used for optical pumping is  $\sigma^+$. }\label{f4}
\end{figure}

\begin{figure}[h!]
\centering
\includegraphics[width=8.5cm]{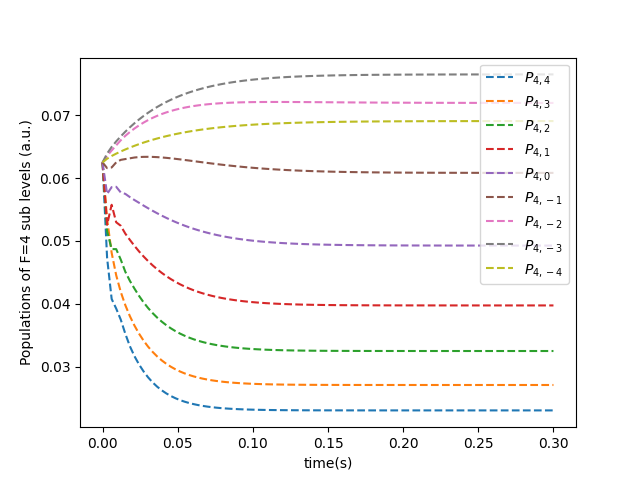}
\caption{Time evolution of the population for the Zeeman sublevels $F=4$. The applied polarized light, used for optical pumping is  $\sigma^-$. }\label{f5}
\end{figure}

Figures~\ref{f4} and~\ref{f5} present the numerical outcomes depicting the time evolution of Zeeman sublevel populations with $F=4$, influenced by the application of polarized light in the $\sigma^+$ and $\sigma^-$ configurations, respectively. The parameters considered include $\Omega_R=11\times 10^3$ Hz, $\Gamma=613$ MHz, $\Omega_L=0.05\times \Gamma$ MHz, $\Delta=0.5\times \Gamma$ MHz, $\gamma=0.05\times \Gamma$ MHz, and $\omega_0=3351.21$ MHz, aligning with experimental findings for Cs atoms~\cite{r23}.

In Fig.~\ref{f4}, the population dynamics of sublevels $F=4$ under the influence of $\sigma^+$ polarized light are illustrated. Notably, the populations of Zeeman sublevels $m_f=4$ and $m_f=3$ exhibit prominent increases compared to other sublevels during the optical pumping process. This population enhancement in specific sublevels can be attributed to the utilization of $\sigma^+$ light in accordance with atomic transition rules~\cite{r24}. Similarly, Fig.~\ref{f5} portrays the population evolution of sublevels $F=4$ driven by $\sigma^-$ light. Here, a noticeable rise in the populations of Zeeman sublevels $m_f=-4$ and $m_f=-3$ is observed relative to other sublevels, emphasizing the impact of using $\sigma^-$ light.

In addition, in Fig.~\ref{f4}, the sequence of sublevel populations is observed as $P_{4,3}>P_{4,2}>P_{4,4}>P_{4,1}$, while in Fig.~\ref{f5}, the order is depicted as $P_{4,-3}>P_{4,-2}>P_{4,-4}>P_{4,-1}$. This disparity in population distribution underscores the distinct effects of employing $\sigma^+$ and $\sigma^-$ light configurations, respectively, attributed to the energy variations among cesium atom sublevels in each scenario.

Furthermore, our investigation delves into the impact of a magnetic-resonance RF field on the temporal evolution of Zeeman sublevel populations. Initially, the atoms are uniformly distributed among the ground-state Zeeman sublevels. Upon absorption of photons from a laser beam during pumping, the atoms transition to Zeeman sublevels within the excited states before undergoing spontaneous decay and returning to the ground-state sublevels. Throughout this optical process, although all atoms reside in the ground-state sublevels, the populations among these sublevels are not uniform. The application of an RF field can counteract this pumping effect; by interacting with all Zeeman sublevels through a relaxation mechanism, the RF field serves to equalize the populations across sublevels~\cite{r24}

In our study, we introduce an RF field to Cs atoms along the $x$ direction with an associated Rabi frequency denoted as $\Omega_{RF}=g\mu_B B_{0RF}$:
\begin{equation}
\begin{array}{ccl}
 B_{RF} = B_{0RF}\cos(\omega_{RF}t),\\
 \hat{H}_{RF} =\boldmath\mu .\hat{\textbf{B}}=\Omega_{RF}\cos(\omega_{RF}t)F_{RF}.
\end{array}
\end{equation}
Here, $F_{RF}$ represents a $32\times 32$ matrix containing elements describing magnetic dipole transitions with $\Delta m=0,\ \pm 1$. Focusing on optical pumping between $F=4$ and $f=3$, we explore magnetic dipole transitions between these specific sublevels. By incorporating the Hamiltonian associated with the RF field into the total Hamiltonian of the system of interest according to Eq.~\ref{eq1}, we apply the rotating wave approximation to the RF field's Hamiltonian, retaining only terms indicative of RF detuning ($\Delta_{RF}=\omega_{RF}-\Omega_L$).

\begin{figure}[h!]
\centering
\includegraphics[width=10cm]{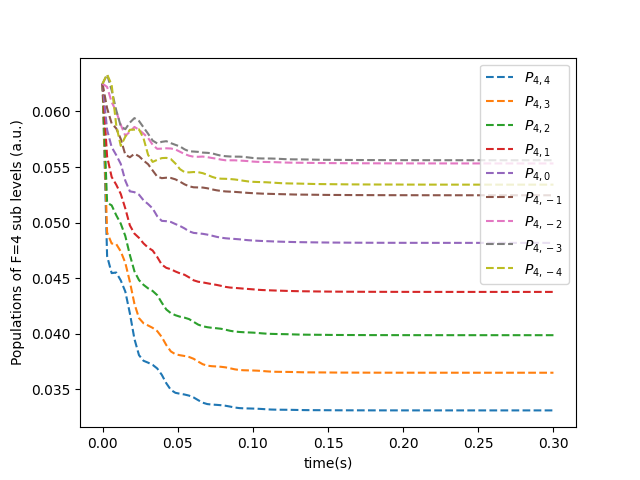}
\caption{Time evolution of the population of Zeeman sublevels $F=4$, for the applied polarized light $\sigma^-$, at the presence of a magnetic-resonance RF field.}\label{f6}
\end{figure}
\begin{figure}[h!]
\centering
\includegraphics[width=10cm]{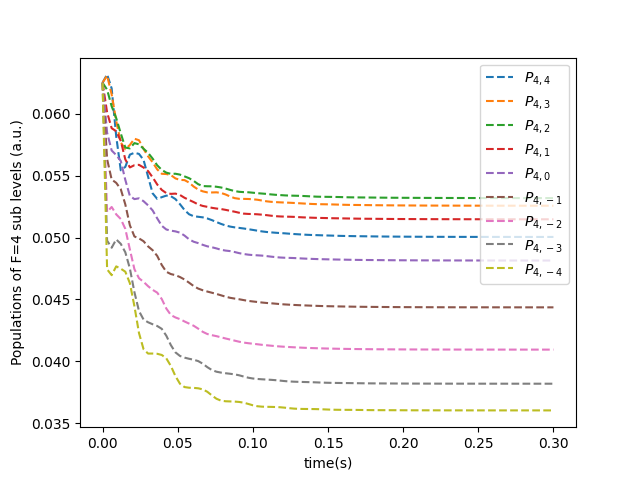}
\caption{Time evolution of the population of Zeeman-sublevels $F=4$, for the applied polarized light $\sigma^+$, at the presence of a magnetic-resonance RF field. }\label{f7}
\end{figure}
Figures~\ref{f6} and \ref{f7} depict the outcomes of incorporating a RF field during the optical pumping process and its consequential impact on the temporal evolution of Zeeman sublevel populations. Figure~\ref{f6} displays the time evolution of Zeeman sublevel populations for $F=4$ under $\sigma^+$ polarized light in the presence of a magnetic-resonance RF field, while Fig.~\ref{f7} showcases the corresponding evolution for $\sigma^-$ polarized light with the RF field applied. For our numerical simulations, we have selected $\Omega_{RF}=2000$ MHz and $\omega_{RF}=\Omega_L$ to align with the resonance condition typically employed in experimental settings~\cite{r23,r25,r24}. In Fig.~\ref{f6}, the sublevel populations of $F=4$ are plotted under the influence of $\sigma^+$ polarized light in the presence of the applied RF field. Notably, as the time steps progress, a convergence towards equal populations among Zeeman sublevels is observed. Similarly, Fig.~\ref{f7} illustrates the equalization of sublevel populations for $F=4$ under $\sigma^-$ light in conjunction with the RF field. 

The application of the RF field effectively equalizes the populations of sublevels in both scenarios. The graphical representation reveals a minimal population difference among sublevels, on the order of 0.01, mirroring experimental observations. Consequently, it can be inferred from the data presented in the figures that the primary function of the RF field is to homogenize the populations across all sublevels.

\begin{figure}[h!]
\centering
\includegraphics[width=10cm]{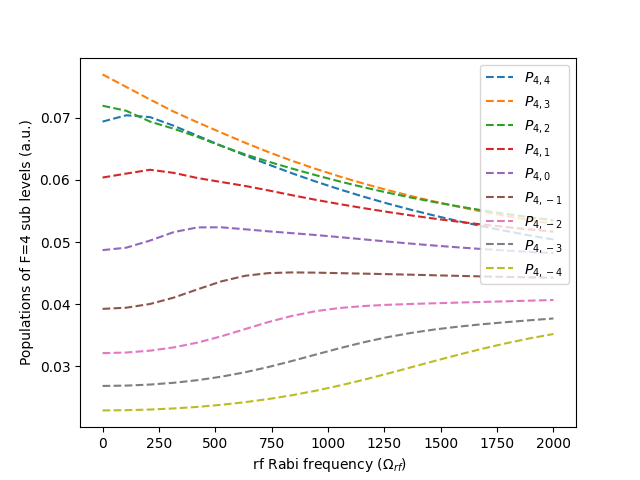}
\caption{Time evolution of the population of Zeeman sublevels $F=4$ for $\sigma^+$ light, as a function of RF Rabi frequency.}\label{f8}
\end{figure}

\begin{figure}[h!]
\centering
\includegraphics[width=10cm]{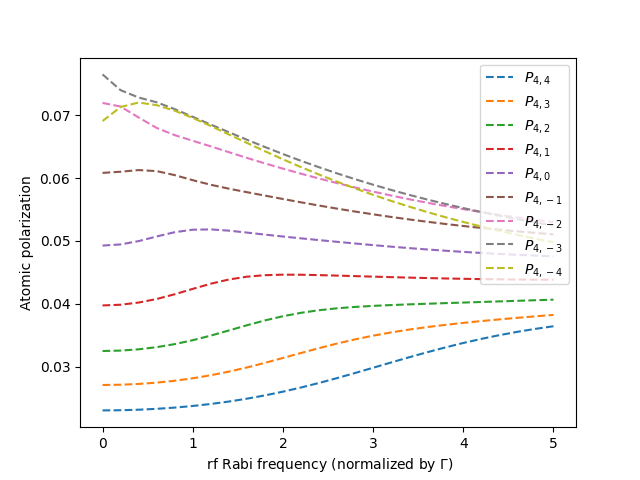}
\caption{Time evolution of the population of Zeeman sublevels $F=4$ for $\sigma^-$ light as a function of RF Rabi frequency.}\label{f9}
\end{figure}

Moreover, we delve into the temporal evolution of Zeeman sublevel populations concerning the Rabi frequency of the RF field. The findings are presented in Figs.~\ref{f8} and~\ref{f9} for the respective cases of applied polarized light, $\sigma^+$ and $\sigma^-$, respectively. In Fig.~\ref{f8}, a notable trend emerges where an increase in the RF frequency leads to a quicker equalization of Zeeman sublevel populations. This behavior is similarly observed in Fig.~\ref{f9}, underscoring the influence of RF Rabi frequency on the homogenization process. Consequently, the results clearly indicate that elevating the Rabi frequency facilitates the equalization of Zeeman sublevel populations.

\begin{figure}[h!]
\centering
\includegraphics[width=10cm]{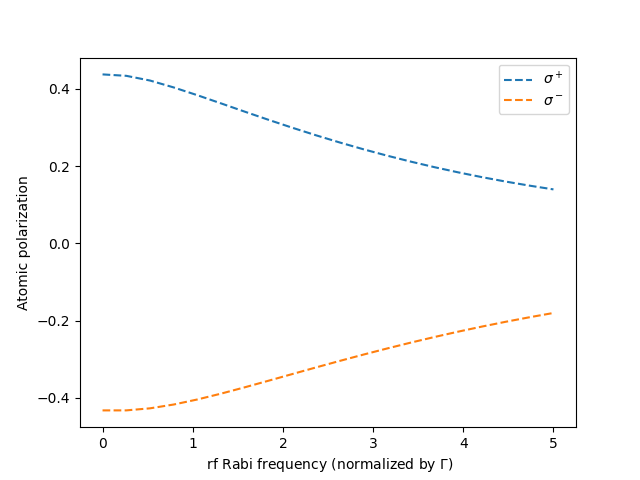}
\caption{Time evolution of the atomic polarization for the applied light $\sigma^+$ and $\sigma^-$.}\label{f10}
\end{figure}

Figure~\ref{f10} delves into the time evolution of atomic polarization concerning $\sigma^+$ and $\sigma^-$ light, in relation to the Rabi frequency of the RF field. Atomic polarization is computed as $\langle F_Z\rangle=1/M \sum_M p_M M$, where $M$ represents the Zeeman sublevel projection number (i.e., different $m_f$) and $p_M$ denotes its population. Notably, as the Rabi frequency increases, the atomic polarization associated with $\sigma^+$ ($\sigma^-$) light demonstrates a decrease (increase). This observation indicates that by inducing an imbalance in Zeeman sublevel populations through the application of an RF field, alterations in polarization can be effectively achieved.

\section{Conclusion}\label{sec4}

In this research endeavor, we delved into the optical pumping process of Cesium atoms utilizing the Liouville equation methodology. Our investigation extended to the manipulation of Cesium Zeeman sublevel populations through the application of a radiofrequency field. Initially, circularly polarized light was directed at Cesium atoms to explore the impact of relaxation and repopulation phenomena. Subsequently, by incorporating the radiofrequency field and its effects within the Liouville equation framework, we successfully engineered the population distribution within Cesium Zeeman sublevels. Through a comprehensive analysis, we examined the temporal evolution of Zeeman sublevel populations under varying conditions, both in the presence and absence of the RF field, during optical pumping with both polarized lights, $\sigma^+$ and $\sigma^-$. Additionally, we scrutinized the time evolution of atomic polarization for $\sigma^+$ and $\sigma^-$ light configurations. This methodological approach holds promise for its applicability across alkali atoms and presents numerous opportunities for diverse applications in a wide array of optical experiments.

\bibliographystyle{unsrt}
\bibliography{re2}

\end{document}